\documentclass[prl,twocolumn]{revtex4}

\usepackage{graphicx}
\usepackage{amsmath}

\graphicspath{{./}{./plots/}}

\newcommand{\dlow}{d_\mathrm{low}}
\newcommand{\dup}{d_\mathrm{up}}
\newcommand{\Tc}{T_\mathrm{c}}
\newcommand{\Nc}{N_\mathrm{c}}
\newcommand{\epsc}{\epsilon_\mathrm{c}}
\newcommand{\eel}{e_\mathrm{el}}
\newcommand{\mel}{m_\mathrm{el}}

\begin{document}

\title{Topological Mott insulator in three-dimensional systems with quadratic band touching}

\author{Igor F. Herbut and Lukas Janssen}

\affiliation{Department of Physics, Simon Fraser University, Burnaby, British Columbia, Canada V5A 1S6}

\begin{abstract} We argue that a three dimensional electronic system with the Fermi level at the quadratic band touching point such as HgTe could be unstable with respect to the spontaneous formation of the (topological) Mott insulator at arbitrary weak long range Coulomb interaction. The mechanism of the instability can be understood as the collision of Abrikosov's non-Fermi liquid (NFL) fixed point with another, quantum critical, fixed point, which  approaches it in the coupling space as the system's dimensionality $d\rightarrow \dlow +$, with the ``lower critical dimension" $2<\dlow<4$. Arguments for the existence of the quantum critical point based on considerations in the large-$N$ limit in $d=3$, as well as close to $d=2$, are given. In the one-loop calculation we find that $\dlow = 3.26$, and thus above, but not far from three dimensions. This translates into a temperature/energy window $(\Tc, T_*)$ over which the NFL scaling should still be observable, before the Mott transition finally takes place at the critical temperature $\Tc \sim T_* \exp(-z C/(\dlow -d)^{1/2})$. We estimate $C=\pi/1.1$, dynamical critical exponent $z\approx 1.8$, and the temperature scale $k_\mathrm{B} T_*\approx (4 m/\mel \varepsilon^2) \,13.6\,\mathrm{eV}$, with $m$ as the band mass and $\varepsilon$ as the dielectric constant.
\end{abstract}

\maketitle

The electronic systems with Fermi points instead of the usual Fermi surface represent a new frontier in quantum condensed matter physics. The most generic among these are the ones with simple band crossing, such as graphene in two dimensions (2D) and Weyl semimetals in three dimensions (3D), which feature emerging relativistic invariance at low energies \cite{vafvish}. The vanishing density of states at the Dirac point in both of these cases protects the semimetallic states against the effects of electron-electron interactions, which need to be above a certain threshold to qualitatively change the ground state \cite{herbjurroy}. From the point of view of electron correlations maybe a more interesting example is the case of non-Dirac quadratic band touching (QBT) \cite{sun}. Important examples in 2D are provided by the bilayer graphene, and by surface states of a topological crystalline insulator \cite{liang}. The former is indeed at zero temperature believed to be unstable towards a broken symmetry phase already at infinitesimal interactions \cite{vafek}. The principal cause for the instability are the short-range interactions, whereas the long-range tail of the Coulomb repulsion between electrons is effectively screened in the presence of the finite density of states  at the touching point, and may essentially be neglected \cite{lemonik}. For a 3D QBT, such as occurs, for example, in gray tin or HgTe \cite{tsidilkovski}, the physics is, on the other hand,  believed to be quite different. The vanishing density of states in 3D now makes the weak short-range components of the Coulomb repulsion irrelevant, but the long-range tail of it is, in turn, not screened. The corresponding coupling constant (hereafter called the ``charge") is a relevant coupling which flows towards a new, interacting, infrared (IR) attractive fixed point, at which the system could become a scale invariant non-Fermi liquid (NFL). This scenario, originally due to Abrikosov \cite{abrikosov}, has recently been revisited and put forward as an explanation of the anomalous low temperature behavior displayed by pyrochlore iridiates \cite{moon}. Besides possibly connecting to experiment \cite{nakatsuji, machida}, the basic result that the 3D electronic system with the chemical potential at the point of QBT is quite generically a NFL in its ground state is certainly of fundamental theoretical importance. In this Letter we give it therefore another critical look, and arrive at conclusions different from those existing in the literature~\cite{abrikosov, moon}.

 The principal result of our analysis is that the Abrikosov's NFL fixed point, besides existing only below the upper (spatial) critical dimension $\dup =4$, survives also only above the {\it lower critical dimension} $\dlow$, with $2<\dlow<4 $. In a fixed dimension $d$, the NFL fixed point survives only for a number $N$ of degenerate QBT points larger than a certain critical value $\Nc (d)$.  This critical number is such that $\Nc(4)<1$ and $\Nc(d\rightarrow 2)\rightarrow \infty$, so by continuity $\Nc(3)$ should be finite.  One loop calculation for the physical case of $N=1$ leads, for example, to $\dlow = 3.26$, and thus {\it larger} than $3$. Directly  in 3D we similarly estimate  $\Nc (3) = 2.07$. The main culprits behind the disappearance of the NFL fixed point are those short range components of the Coulomb interaction, which inevitably become generated during Wilson's renormalization group (RG) procedure, and which one is tempted to discard as irrelevant. We argue that one of these generated short range couplings is particularly dangerous; if large enough, it would cause an opening of the  gap in the spectrum, and a transition into a phase with broken rotational invariance. Such an insulating phase, in materials with the band structure equivalent to that of gray tin for example, at the mean-field  level could be identified as a strong topological insulator \cite{fukane}.

 At zero charge there are therefore {\it two} fixed points: the (IR stable) Gaussian fixed point (G), and the (ultraviolet (UV) stable) strong coupling quantum critical point (QCP$_0$) (Fig.~1). As the value of the charge at the  IR stable fixed point grows with the increase of the parameter $\epsilon=4-d$ or by decrease in $N$, these two fixed points approach each other, and ultimately collide in the coupling constant space at some critical value (Fig.~2). Beyond this critical value of the charge, at a large $\epsilon$ or small $N$, both fixed points turn complex, leaving only a runaway flow in the physical space of real couplings. Such a runaway flow is identified as an instability towards the Mott insulator. Finally, we show that if $\dlow$  is above but close to the physical dimension $d=3$,  there will be a relatively wide range of temperature or energy over which the scale invariance of the NFL state should still be manifest. The Mott gap or the transition temperature are estimated to be $\sim 1\,\mathrm{K}$ in gray tin or HgTe.  The topological Mott insulator would be expected to display an  anisotropic transport, the appearance of the Dirac-like surface states, and the concomitant characteristic quantum Hall effect. There should also be a thermodynamic singularity marking the finite temperature transition, evident in the specific heat, for example.

 \begin{figure}
\includegraphics[width=0.33\textwidth]{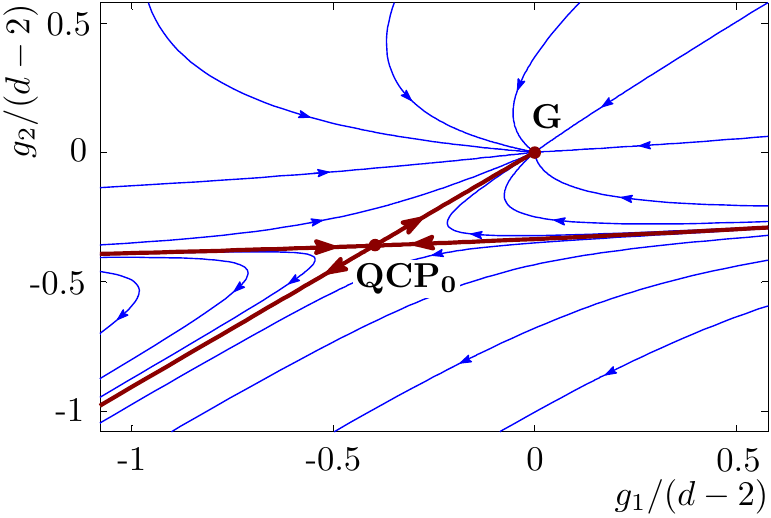}
\caption{The quantum critical (QCP$_0$) and the Gaussian (G) fixed points in the flow diagram at zero charge.}
\end{figure}

\paragraph{Lagrangian and its symmetries.} We are interested in the 3D system of interacting fermions with the Euclidian action $S=\int d\tau d\vec{x} L$, and with the Lagrangian
\begin{align}
L & = \Psi^\dagger \left(\partial_\tau + i a + d_i (-i\nabla) \gamma_i\right)\Psi +  \nonumber \\
&\quad
g_1 (\Psi^\dagger \Psi)^2 + g_2 (\Psi^\dagger \gamma_i \Psi)^2  + \frac{1}{2e^2} (\nabla a)^2.
\end{align}
The summation convention is assumed, $\Psi$ is a four-component Grassmann field, and the matrices $\gamma_i$, $i=1,\dots,5$ provide a four-dimensional (Hermitian) representation of the Clifford algebra $\{ \gamma_i, \gamma_j \}=2\delta_{ij}$. $ d_i (\vec{p})=p^2 \tilde{d}_i (\theta,\phi)$ are proportional to five spherical harmonics for the angular momentum of {\it two}; explicitly, $\tilde{d}_1 + i \tilde{d}_2 = (\sqrt{3}/2)\sin^2 (\theta) e^{2i\phi}$,  $\tilde{d}_3 + i \tilde{d}_4 = (\sqrt{3}/2)\sin (2\theta) e^{i\phi}$, $\tilde{d}_5= (3 \cos^2 \theta -1)/2$, with $\theta$ and $\phi$ as the spherical angles in the momentum space \cite{luttinger, murakami}. Their form assures that the energy spectrum of the single particle Hamiltonian $H_0= d_i(\vec{p}) \gamma_i$ featured in $L$ is simply $E=\pm p ^2$, and doubly degenerate at all momenta.

The scalar field $a$ mediates the long-range $\sim 1/p^2$ density-density interaction, or, in real space, $\sim 1/r^{d-2}$. The contact interactions $g_1$ and $g_2$ will be generated by the change of the UV cutoff $\Lambda$, and we include them therefore from the outset.\cite{z} Note that the forms of all interaction terms are invariant under the spinor representation of the group $SO(5)$, generated by the ten generators $i \gamma_i \gamma_j$, $i>j$. The kinetic energy term, however, is invariant only under the real space $O(3)$ rotations.

The Hamiltonian $H_0$ is nothing but the standard isotropic Luttinger Hamiltonian: $H_0 = ((5p^2/4)-(\vec{p}\cdot \vec{J})^2)/(2m)$, with $\vec{J}$ as the ``angular momentum" 3/2 generators, and with the band mass set to $2m=1$ \cite{abrikosov, moon, murakami}. The form of the Luttinger Hamiltonian is dictated by the dimensionality of four of the representation of the crystal's cubic symmetry, and the $k \cdot p$ theory near the $\Gamma$ point \cite{luttinger}. To keep the discussion simple, we have omitted the remaining symmetry-allowed terms, proportional to the unit matrix in the Hamiltonian, which would introduce particle-hole asymmetry, and the cubic anisotropy \cite{luttinger, dora}. These are also known to be irrelevant in the presence of Coulomb interaction \cite{abrikosov, moon}. The dielectric constant of the host material $\varepsilon$, the band mass $m$, and the Planck's constant $\hbar$ have been absorbed into the charge, which is then $e^2= 2m \eel^2 /( 4 \pi \hbar ^2 \varepsilon)$. $\eel$ is the electron's charge.

The Lagrangian $L$ is also invariant under the transformation $\Psi\rightarrow A \Psi$, with the {\it unique} antilinear operator $A$. Choosing $\gamma_i$ to be real for $i=1,2,3$ and imaginary for $i=4,5$, for example \cite{herbutTR}, yields $A = \gamma_4 \gamma_5 K$. It is then evident that $A^2=-1$, and the four-component QBT must originate from the spin-orbit coupling, since it describes a particle with a half-integer spin. The six fermion bilinears $\Psi^\dagger \Psi$ and $\Psi^\dagger \gamma_i \Psi$ are then all {\it even} under the time reversal, whereas all ten of $i \Psi^\dagger \gamma_i \gamma_j \Psi$ are {\it odd}.

\begin{figure*}
\includegraphics[width=0.33\textwidth]{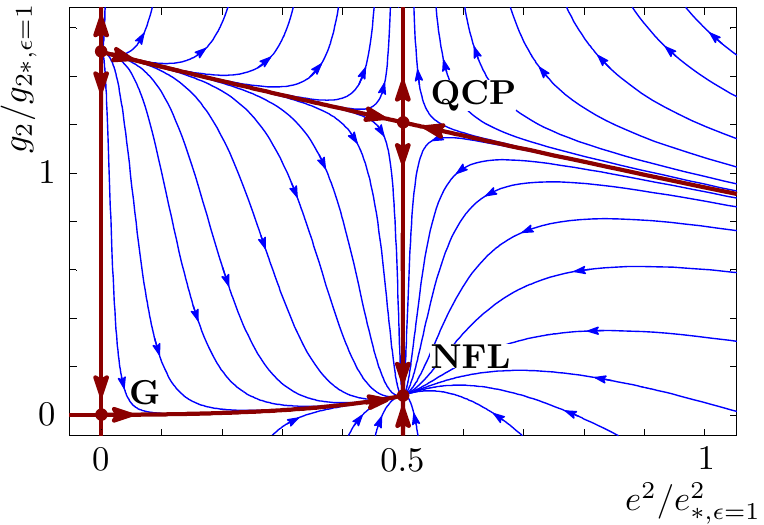}\hfill
\includegraphics[width=0.33\textwidth]{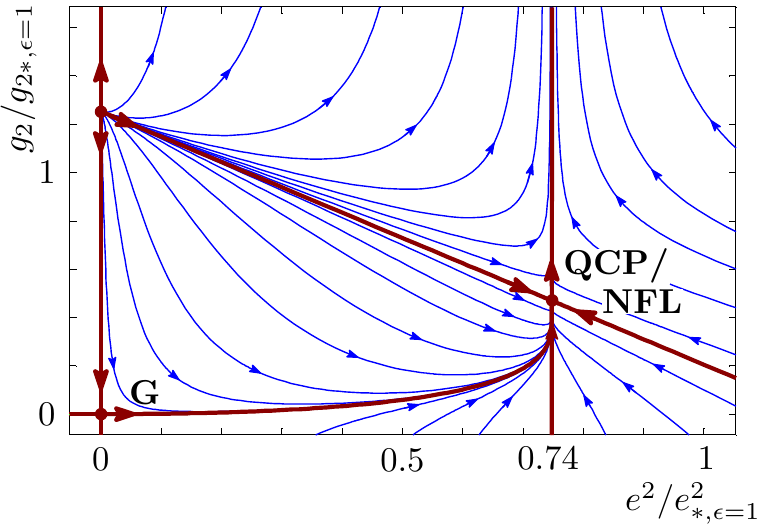}\hfill
\includegraphics[width=0.33\textwidth]{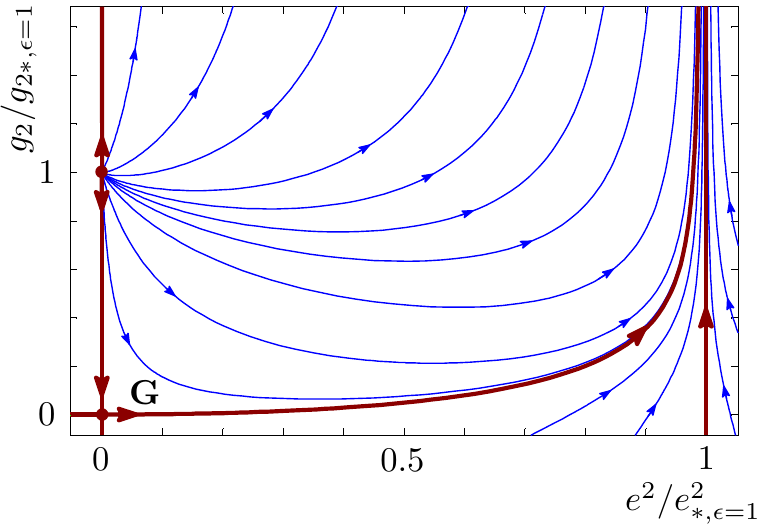}
\caption{The flows in $e^2$-$g_2$ plane for different spatial dimensions $d$: $\dlow<d<\dup$ (left, $d=3.5$), $d=\dlow$ (middle, $d=3.26$), and $d<\dlow$ (right panel, $d=3$). ($g_1$ has been neglected as qualitatively and quantitatively unimportant for $d\leq 3.5$.) }
\end{figure*}

\paragraph{Renormalization.} Integrating over the fermionic modes with the momenta from $\Lambda/b$ to $\Lambda$ and with all frequencies at $T=0$ \cite{book} causes the coupling constants to flow according to the equations,
\begin{equation}
   \frac{dg_1}{d\ln b} = (z-d) g_1 -\frac{(g_1 + e^2) g_2}{2} - \frac{5g_2 ^2}{2},
\end{equation}
\begin{equation}
   \frac{dg_2}{d\ln b} = (z-d) g_2 + \frac{2(g_1+  e^2) g_2}{5} - \frac{(g_1 +  e^2) ^2}{20}  - \frac{ 63 g_2 ^2}{20},
\end{equation}
\begin{equation}
   \frac{de^2 }{d\ln b} = (z+2 -d) e^2  -  \frac{e^4}{2},
\end{equation}
after the rescaling of the couplings as $2 g_i \Lambda^{d-z} /\pi^2  \rightarrow g_i$ and $e^2 \Lambda^{d-z-2} /  \pi^2 \rightarrow e^2$, to the leading order. The angular integrals are performed in $d=3$, and the dimensions of the couplings are counted in general $d$. The dynamical critical exponent is $z= 2-(2 e^2/15) +O(e^4)$.

Eq. (4) for the flow of the charge is equivalent to the previous result \cite{moon}. For $d>4$ the charge $e^2 $ is IR irrelevant at the Gaussian fixed point, whereas for $d<4$ it is relevant and attracted to the new fixed point.  $d=4$ is thus the upper critical dimension. Eq.~(3) implies that the (negative) couplings $g_1$ and $g_2$ become generated by the charge, even if absent initially. We have eliminated the third $SO(5)$-symmetric coupling that arises by using the Fierz identity \cite{herbjurroy} $(\Psi^\dagger \gamma_i \gamma_j \Psi)^2= 15 (\Psi^\dagger \Psi)^2 + 2 (\Psi^\dagger \gamma_i \Psi)^2 $. The space $(g_1, g_2, e^2)$  is then closed under the RG to the leading order.

\paragraph{Quantum critical point at $e=0$.}  Besides the IR stable Gaussian fixed point, in any dimension $4\geq d >2$ and at $e=0$ there is also a (UV stable) QCP at negative values of both $g_1$ and $g_2$ (Fig.~1). QCP  approaches the Gaussian fixed point as $d\rightarrow 2+$, but in $d=3$ it is still located at strong coupling. In order to gain more confidence in its existence, we consider the Lagrangian in Eq.~(1) with $e^2=g_1 =0$. Adding a flavor index to fermions as $\Psi\rightarrow \Psi_\alpha$, $\alpha=1,..., N$ and then summing over it \cite{remark1}, the flow equation for $g_2$ in $d=3$ and for large $N$ would be simply \cite{remark2}
\begin{eqnarray}
   \frac{dg_2}{d\ln b} = -g_2   - \frac{4N }{5} g_2 ^2.
\end{eqnarray}
 In the large-$N$ limit, on the other hand, the theory is exactly solved by the saddle-point method, which yields the {\it order parameter} $\chi_i = 2 g_2 \langle \Psi^\dagger \gamma_i \Psi \rangle$ self-consistently as
\begin{equation}
\chi_i = -4 g_2 N \int \frac{d\vec{p}}{ (2\pi)^3 } \frac{ d_i (\vec{p}) + \chi_i}{ \sqrt{ (d_j (\vec{p}) + \chi_j)^2   } }.
\end{equation}
The straightforward analysis of Eq. (6) shows that the stable saddle point becomes non-trivial for $g_2 < g_c$, where
\begin{equation}
g_c ^{-1} = 4 N \int  \frac{d\vec{p}}{ (2\pi)^3 } \frac{ d_i ^2  (\vec{p}) - p^4}{p^6} = - \frac{4 N}{5} \frac{ 2\Lambda}{ \pi ^2},
\end{equation}
{\it without} the summation convention in the last line. Note that the critical coupling is independent of the ``direction" in the 5-dimensional space of the components $d_i (\vec{p})$. The zero of the flow equation for $g_2$ in the large-$N$ limit corresponds to the exact critical coupling. Since the quantum phase transition should persist at all $N$, we expect that the critical point at $e^2=0$, although located at strong coupling in $d=3$, is indeed a genuine feature of the theory, and present even when $N=1$.

To understand the ensuing ordering we examine the spectrum when some of the order parameters $\chi_i\neq 0$. For example, if $\chi_1 \neq 0$ (and $\chi_{i\geq 2} = 0$),
  \begin{equation}
  E^2 = p^4 + \chi_1 ^2 + 2 \chi_1 d_1 (\vec{p}) \geq p^4 + \chi_1 ^2 - \sqrt{3} p^2 |\chi_1|,
  \end{equation}
  with the minimal gap of $|\chi_1|/2$ being located at $p^2 = \sqrt{3} |\chi_1|/2$, $\theta=\pi/2$, and $\phi= \pi/2, 3\pi/2$ (for $\chi_1 >0$, for example). Similar spectrum with the minimal gap at two opposite points in the momentum space follows for every individual $\chi_i$ with $i=1,2,3,4$. If $\chi_5>0$, on the other hand, the minimal gap of $E_0 = \sqrt{3} \chi_5 /2$ is at the entire {\it circle} located at $p^2 = \chi_5/2$ and $\theta= \pi/2$. Finally, the spectrum is {\it gapless} if $\chi_5 <0$, with the gap closing at two points at $p^2 = |\chi_5|$ and $\theta= 0,\pi$.

In all the cases a finite order parameter $\chi_i$ implies breaking of the $O(3)$ rotational symmetry. Although all five order parameters have the same critical coupling, a detailed analysis \cite{janssen} shows that the energy is minimized by developing a positive $\chi_5$, which leads to the largest gap in the spectrum. The mean-field Hamiltonian $H_\mathrm{MF} = H_0 + \chi_5 \gamma_5$, on the other hand, can be recognized as describing the system under strain that preserves a rotational symmetry around $z$-axis \cite{moon, roman}. Of course, the choice of $z$-axis is completely arbitrary; since order parameters $\chi_i$ transform as $l=2$ spherical harmonics under spatial rotations, the same order parameter with a different axis of symmetry $z'$ would be $\chi_i ' = \chi_5 \tilde{d}_i (\theta',\phi')$, with $(\theta', \phi')$ as the polar angles of the axis $z'$.

The ground state of $H_\mathrm{MF}$ for $\chi_5 >0$ is therefore an insulator with a gap. Taking into account the complete band structure that contained the QBT at the $\Gamma$ point in gray tin for example, and computing the parity eigenvalues of the occupied states at the time-reversal-invariant crystal momenta leads to the conclusion that the $Z_2$ topological invariant of such a gapped state is non-trivial \cite{fukane}. In this case the gapped ground state which  results from the Coulomb interaction may be understood as the dynamically generated strong topological {\it Mott} insulator.

\paragraph{Finite charge.} Finally, consider $e\neq 0$, and $d<4$. Introducing the parameter $\epsilon = 4-d$ one can control the magnitude of the charge at the IR stable fixed point: $e_* ^2 = 30 \epsilon/19+ \mathcal{O}(\epsilon^2)$. One then finds the Gaussian and the QCP getting closer in the $g_1$-$g_2$ plane, but remaining real and separate as long as  $\epsilon < \epsc$, with $\epsc = 0.74$. The stable NFL fixed point at finite charge descends from the Gaussian fixed point (Fig.~2, left panel) and has a finite domain of attraction  for dimensions $\dup>d>\dlow$, with the {\it lower critical dimension} $\dlow = 4-\epsc=3.26$. The NFL state is scale invariant, with characteristic power-law response functions, and with $z\neq 2$, for example  \cite{abrikosov,moon}. For too negative $g_1$ or $g_2$ and $\dup > d>\dlow$ the system spontaneously breaks the rotational symmetry by developing  a positive $\chi_5$. For $d<\dlow$, however, the flow is qualitatively different, and there are no  charged fixed points left  (Fig.~2, right panel).   Consequently, the flow is always towards the region with large negative short range couplings. The reason for this is physical, and clear from Eq.~(3): the charge generates a {\it negative} $g_2$, and thus favors the development of $\chi_5$. In fact, neglecting completely $g_1$ leads to qualitatively identical, and even quantitatively very similar picture, in which, for example, $\epsc=19 (3-\sqrt{7})/9 =  0.75$.

 Generalizing to $N>1$ requires some care with Fierz identities, and we only quote here the main results \cite{janssen}. As evident already in Eqs.~(2)--(3), the critical point QCP$_0$ at $e=0$ becomes weakly coupled  as $d\rightarrow 2+$. Infinitesimal charge is therefore enough to collide the NFL and the QCP near $d=2$. This way one finds that $\Nc (d) \sim 1/(d-2)^2$, and the possibility of the NFL fixed point disappearing completely as $d\rightarrow 2$. Since on the other hand $\Nc(4) <1$, $\Nc(3)$ is finite by continuity. We find from the generalization of the Eqs.~(2)--(4) to finite $N$ the value $\Nc(3) = 2.07$. In accord with our calculation of $\dlow$, this estimate places the physical case of $d=3$ and $N=1$ on the insulating side, but close to the phase boundary in the $(d,N)$ plane.

\paragraph{Hierarchy  of scales.} Although in $d=3$ the Eqs. (2)--(4) imply that the ground state is an insulator with broken rotational symmetry, the system may still be influenced by the complex fixed point, if its imaginary part is small. By integrating the flow equations it is easy to show that as $d \rightarrow \dlow - $ the RG ``time" $b$  it takes the short range couplings to diverge is
   \begin{equation}
   b_0 = e^{ \frac{C}{\sqrt{\dlow-d}} -B + \mathcal{O}(\dlow-d) },
   \end{equation}
with non-universal constants $C=\pi/1.1$ and $B=2.1$. This means that close to and below the lower critical dimension the broken-symmetry state will reveal itself only at the energies (or temperatures) typically much smaller than the cutoff, leaving possibly a significant window of temperatures where the NFL scaling in various physical quantities \cite{moon} may still be observable. Such separation of scales is a generic consequence of the complexification of the fixed points under the change of some parameter. Other examples include the fluctuation-induced first order transition in superconductors \cite{HLM, zlatko1, zlatko2, book}, chiral symmetry breaking in three dimensional quantum electrodynamics \cite{kaveh, braun} and in quantum chromodynamics \cite{gies, kaplan}.

\paragraph{Time reversal symmetry breaking?} The reader may object to our elimination of the term $(\Psi ^\dagger \gamma_i \gamma_j \Psi)^2$  in favor of the other two, particularly since the related fermion bilinear $\langle \Psi ^\dagger i \gamma_i \gamma_j \Psi \rangle$ that violates {\it both} the time reversal and the rotational symmetry would be expected to become finite in the theory with a term $g_3 (\Psi ^\dagger i \gamma_i \gamma_j \Psi)^2 $, $i\neq j$ \cite{savary}.  Assuming all short-range couplings to be weak and allowing them to be generated by the charge, however, the runaway flow is always towards large and negative $g_1$ and $g_2$. If one of these is traded for the coupling $g_3$, $g_3$ would flow towards large, but {\it positive} values. It is easy to check, on the other hand, that only a large negative $g_3$ (as defined above) would lead to spontaneous breaking of the time reversal symmetry.

\paragraph{Experimental relevance.} We can estimate the relevant energy (temperature) scales for the emergence of the NFL scaling, as well as for the ultimate Mott  transition. Recalling the definition of the charge, the crossover length scale at which the NFL behavior sets in is $L_* \sim 1/(4 \pi e^2$). Put differently, the corresponding energy scale is
 \begin{equation}
k_\mathrm{B} T_* \sim \frac{\eel^2}{\varepsilon L_*} = \frac{\hbar^2} {2 m L_* ^2} = \frac{4m}{\mel \varepsilon^2 } E_0,
\end{equation}
where $\mel$ is the mass of electron, and $E_0 = 13.6\,\mathrm{eV}$. Assuming  a band mass as low as  $m/\mel\approx 1/50$ and high dielectric constant $\varepsilon\approx 30$ leads to the crossover temperature $T_* \sim 10\,\mathrm{K} - 100\,\mathrm{K}$. The critical temperature of the Mott transition is then
\begin{equation}
\Tc \approx T_* b_0^{-z},
\end{equation}
so that $\Tc \approx T_* / 100$, assuming $z\approx 1.8$.

Both $T_*$ and $\Tc$ do not appear to be unobservable in gapless semiconductors such as grey tin or HgTe, if prepared sufficiently pure \cite{molenkamp}. Doping the system with electrons or holes introduces a new length scale, which cuts off the flows we exhibited. If large enough, doping will restore the usual Fermi liquid. The average separation between the doped carriers needs to be longer than $L_*$ for the interaction effects discussed here to be relevant.

\paragraph{Conclusion.} We presented arguments that the tail of the Coulomb  interaction in a 3D electronic system with the chemical potential at the quadratic touching point may lead to the spontaneous breaking of the rotational invariance, and the formation of the (topological) Mott insulator at low temperatures. The mechanism of the instability is the collision of Abrikosov's NFL fixed point with the QCP as the dimensionality of the system approaches the {\it lower critical dimension} $\dlow$ from above. $\dlow$ is estimated to be above and close to three. With lowering of the temperature the system such as very pure HgTe should first display a crossover region of an effective NFL scaling, before the scale invariance is cut off at the critical temperature at which an anisotropic Mott gap opens.

 The authors are grateful to F. Assaad, L. Classen, B. D\' ora, L. Fritz, H. Gies, P. Goswami, Y. B. Kim, L. Molenkamp, C. Mudry, B. Roy, and O. Vafek for useful discussions. This work was supported by the NSERC of Canada (IFH), and the DFG under JA 2306/1-1 (LJ).


\begin{thebibliography}{99}
\bibitem{remark2} In Eq.~(3) the coefficient $63/20$ of the $g_2 ^2$ term is in fact $(47 +16 N)/20$.
\end{thebibliography}
\end{document}